\documentclass{PoS}

\usepackage[alsoload=hep]{siunitx}

\DeclareSIUnit\eVperc{\eV\per\clight}
\newcommand{\perc}{\%}

\title{The Run Control system of the NA62 experiment at CERN SPS}

\ShortTitle{The Run Control system of the NA62 experiment at CERN SPS}

\author{\speaker{Cristina Lazzeroni} On behalf of the NA62 collaboration
		\thanks{Participating Institutes: Birmingham,  Bratislava, Bristol, Bucharest, CERN, Dubna, Fairfax, Ferrara, Florence, Frascati, Glasgow, Liverpool, Louvain, Mainz, Merced, Moscow, Naples, Perugia, Pisa, Prague,
Protvino, Rome I, Rome II, San Luis Potosí, Stanford, Sofia, Turin}\\
        School of Physics and Astronomy, University of Birmingham, Birmingham B15 2TT, UK\\
        E-mail: \email{c.lazzeroni@bham.ac.uk}}

\author{Nicolas Lurkin\\
        School of Physics and Astronomy, University of Birmingham, Birmingham B15 2TT, UK\\
        E-mail: \email{nicolas.lurkin@cern.ch}}

\abstract{
	The NA62 experiment at CERN SPS started physics data taking this year with the aim to measure the ultra-rare decay $K^+\to\pi^+\nu\bar{\nu}$.
	The experiment consists of a large number of subsystems to reach the goal of \SI{10}{\perc} accuracy and less than \SI{10}{\perc} background.
	The Run Control has been designed to link their trigger and data acquisition system in a single central application easily controllable by non-expert shifters.
	It has been continuously evolving over time, integrating new equipments, following new requirements and feedback from shifters.
	The next steps of development is a more automatized system that integrates the knowledge acquired during the operation of the experiment.
}

\FullConference{38th International Conference on High Energy Physics\\
		3-10 August 2016\\
		Chicago, USA}

\begin{document}

\section{Introduction}
The NA62 experiment at the CERN SPS aims to collect about 50 events per year of the ultra-rare kaon decay $K^+\to\pi^+\nu\bar{\nu}$ with less than \SI{10}{\perc} background and \SI{10}{\perc} accuracy.
This process is forbidden at tree level and can occur only through a Flavour Changing Neutral Current loop. 
It is theoretically very clean, the short distance contribution dominates and the hadronic matrix element can be related to other well measured processes. 
The prediction of the branching fraction by the Standard Model has been computed to an exceptionally high degree of precision \cite{Buras_2015}: ${\mathcal{B}(K^+\to\pi^+\nu\bar{\nu})_\text{th} = \num{9.11(72)e-11}}$. 

This channel is therefore an excellent probe of the standard model (SM) as any small deviation from this value would indicate new physics.
A total of 7 events were detected at the  E787/E949 experiments at Brookhaven National Laboratory \cite{E787/E949}.
The measured branching ratio ${\mathcal{B}(K^+\to\pi^+\nu\bar{\nu})_\text{exp}=(17.3^{+11.5}_{-10.5})\times 10^{-11}}$ has an uncertainty of about \perc{60} which does not allow to conclude on any deviation from the SM predictions. 

To achieve this ambitious goal, NA62 needs to consider at least $10^{13}~K^+$ decays from the secondary 75 GeV/c unseparated hadron beam with $\sim$6\% $K^+$. 
The in-flight decay technique sets the scale of the detector which has a longitudinal extension exceeding 200~m, enclosing the 65~m long fiducial decay region.

\section{The NA62 Run Control}
The Run Control is centrally controlling and monitoring all the equipment involved in the trigger and data acquisition (TDAQ) system.
Its purpose is to allow a non-expert shifter to supervise the data taking easily while still giving specialists the possibility to achieve a high level of control of their own equipment.

\subsection{Technologies and architecture}
In order to have a coherent system along with the other central control systems of the experiment (DCS, Gas System, Cryogenics) the industrial software ``WinCC Open Architecture" was a natural choice. 
This software is the central part of two frameworks - JCOP \cite{JCOP} and UNICOS \cite{UNICOS}- already widely used at CERN for the LHC experiments. 
Some of the components provided by these developer toolkits are of special relevance for NA62:
\begin{itemize}
  \item DIM (Distributed Interface Management) as the communication layer between the Run Control and the equipment spread across the experiment \cite{DIM}.
  \item The FSM toolkit managing finite state machines with SMI++ processes \cite{SMI++} according to their definition (states, transition rules, actions).
  \item The configuration database tool to define and apply sets of parameters from database.
  \item The Farm Monitoring and Control for the management of the PC farm.
\end{itemize}

The TDAQ system of the experiment is composed of numbers of devices, each of them operating differently.
This complication is overcome by internally using a Finite State Machine (FSM) model of the device, presenting a simple common control to the shifter.
Little equipment specific knowledge is integrated in the Run Control, instead a common generic device interface is created to transmit commands and information.
Simple basic instructions are sent through this channel, and are received by the device control software implementing the standardized interface and encapsulating the knowledge of the device.
It is responsible for executing the proper sequence of actions on the hardware.
More complex configuration relies on a scheme of flexible XML files where the details of the configuration are again handled by the receiver.
Multiple sets of configurations are stored in an Oracle database and can be quickly loaded and distributed to the relevant equipment between runs.
This architecture allows the Run Control and the controlled systems to evolve independently from each other while keeping a constant compatibility. 

The software is implemented as a hierarchical tree of FSM where the state of each node is either defined by a set of rules summarizing the state of the children nodes or by the evaluation of logical expressions involving parameters transmitted by the devices.

The external nodes of the tree implement the FSM models representing the specific hardware or software devices (boards, computer, crates and control software programs). 
Their states are computed according to the value of parameters received from the corresponding piece of equipment. 
The internal nodes represent logical subsystems (subdetectors, PC farm), summarizing the states of their own children nodes following a set of rules. 
Finally the top node further aggregates the state of the subsystems to represent the global state of the experiment.

A change of state is always propagated upwards from the device where this change originated to the top node. 
Conversely, a command can be issued at any node and is always propagated downwards to all the children until it reaches a device node. 
At this point, the command is generated and transmitted to the standardized interface through the network using the DIM protocol. 

\subsection{Infrastructure}
The Run Control is a distributed system spread across different machines in the experiment. 
The core of the system is located on a dedicated WinCC OA data server on the technical network. 
A machine bridging the networks hosts the DIM managers which are responsible for loosely binding all the devices in the experiment and the Run Control. 
The DAQ of the experiment will continue to run correctly by itself in case of a connectivity problem or if this node is cut from one or the other network. 
As soon as communication is re-established, all the clients will reconnect to the DIM servers and the Run Control will resume control over the experiment. 
The bridge also receives external information such as the start and end of burst signals or information about the current run from the accelerator and beam line. 
The user interface itself runs on a different computer in the control room and is remotely connected to the main system.

\subsection{Future developments}
The first version of the Run Control was succesfully deployed for a dry run in 2012 and continuously evolved since then.
New equipment and subdetectors have been delivered, that needed to be integrated.
The experience acquired during the following runs and the feedback from the shifters matured the Run Control to a good level of reliability and usability.

The next steps in the development are the automatization of known procedures and automatic error detection and recovery.
The ELectronic Eye of NA62 (ELENA62) is the core system that is being developed, and already partially deployed, to achieve this goal. 
It provides a framework for monitoring functions, interactions with the shifter through visual and audio notifications, and confirmation windows.
It is highly customizable and configurable as the monitoring and control of specific components of the experiment are provided through plug-in modules.

Few have already been developed and are in permanent use to increase the efficiency and quality of the data taking.
The PC farm module handles the pc nodes by restarting the acquisition software after a crash is detected, or reboots the nodes when a hardware error occurs.
The magnets of the beam line and the vacuum in the decay region are not under supervision of the Run Control but can significantly impact the data quality.
Two ELENA62 modules are implemented to monitor them and provide a fast feedback to the shifters for a quick reaction time in case of problem.
Another module helps the shifter by automatizing the start and end of run procedure, additionally guiding him when a manual input is necessary.

\section{Conclusion}
The NA62 Run Control has already proved its usability and good reliability, confirming the technological choices made for its design.
It has now reached a level of development were day to day maintenance is minimal, allowing the integration of the experience acquired during its operation into a more autonomous system.

\end{document}